\pdfoutput=1

\documentclass[10pt,twocolumn,reprint,amsmath,amssymb,amsfonts,aps,prd,superscriptaddress,showpacs,nofootinbib]{revtex4-2}
\usepackage{graphicx}
\usepackage{hyperref}
\usepackage[normalem]{ulem}
\usepackage{slashed}
\usepackage{float}
\usepackage[caption=false]{subfig}
\usepackage{bm}

\hypersetup{
	bookmarks=true,         
	unicode=true,          
	pdftoolbar=true,        
	pdfmenubar=true,        
	pdffitwindow=false,     
	pdfstartview={FitH},    
	pdftitle={PADMEX17},    
	pdfauthor={Author},     
	pdfsubject={Subject},   
	pdfcreator={Creator},   
	pdfproducer={Producer}, 
	pdfkeywords={keyword1} {key2} {key3}, 
	pdfnewwindow=true,      
	colorlinks=true,       
	linkcolor=black,          
	citecolor=black,        
	filecolor=black,      
	urlcolor=black           
}
 \usepackage[dvipsnames]{xcolor}

\definecolor{nicered}{rgb}{0.62,0.07,0.07}
\definecolor{nicegreen}{rgb}{0.1,0.5,0.1}
\definecolor{red}{rgb}{1.0, 0, 0}
\definecolor{darkblue}{rgb}{.0,.0,.8}
\definecolor{niceblue}{rgb}{0,0,1}
\definecolor{niceviolet}{rgb}{0.5,0,1.0}
\definecolor{blue}{rgb}{0,0,1}
\hypersetup{colorlinks=true,citecolor=nicegreen,linkcolor=nicered,urlcolor=nicered}

\newcommand{\eqn}[1]{Eq.~(\ref{#1})}

\newcommand{\calchep}{\texttt{CalcHEP} }

\def\xsv{X_{17}}
\def\mxsv{M_X}
\def\Eres{E_{\rm  res}}

\newcommand{\gv}[1]{g_{v\!\, #1}}
\newcommand{\gvt}[1]{\tilde{g}_{v\!\,#1}}
\newcommand{\ga}[1]{g_{a\!\,#1}}

\newcommand*{\ROMA}{Dipartimento di Fisica, Universit\`a di Roma La Sapienza and INFN, Sezione di Roma, I-00185 Rome, Italy}	
\newcommand*{\INFNFR}{Istituto Nazionale di Fisica Nucleare, Laboratori Nazionali di Frascati, C.P. 13, 00044 Frascati, Italy}
\newcommand*{\IPII}{Institut de Physique des 2 Infinis de Lyon (IP2I),
UMR5822, CNRS/IN2P3, F-69622 Villeurbanne Cedex, France}

\newcommand*{\ROMATV}{Dip. Fisica, Università di Roma “Tor Vergata”, and INFN sezione di Roma “Tor Vergata”, 00133 Rome, Italy }		
\begin{document}

\title{\bfseries Resonant search for the X17 boson at PADME}


\author{\vspace{0.5cm} Luc Darm\'e}
\email{l.darme@ip2i.in2p3.fr}
\affiliation{\IPII}

\author{\vspace{0.5cm} Marco Mancini}
\email{mancinima@lnf.infn.it}
\affiliation{\ROMATV}	

\author{\vspace{0.5cm} Enrico  Nardi}
\email{enrico.nardi@lnf.infn.it}
\affiliation{\INFNFR}

\author{\vspace{0.5cm} Mauro Raggi}
\email{mauro.raggi@roma1.infn.it}
\affiliation{\ROMA}

\begin{abstract}
We discuss the experimental reach of the Frascati PADME experiment 
in searching for new light bosons via their resonant production 
in positron annihilation on fixed target atomic electrons.
A scan in the mass range around 17\,MeV will thoroughly  
probe  the particle physics interpretation of 
the anomaly observed by the ATOMKI nuclear physics experiment. In  
particular, for the case of a spin-1 boson, the viable 
parameter space can be fully covered in a few months of data taking.
 \end{abstract}

\maketitle


\newpage
\section{Introduction}
\label{intro}

Nuclear excited states with typical energies up to  
several MeV can source in their transition to the ground 
state new light bosons with  MeV masses.
Searches for  signals of new physics (NP) of this type are 
carried out for example at the ATOMKI Institute for Nuclear Research 
in Debrecen (HU), that recently 
reported an anomaly in the angular correlation spectra in $^8$Be and $^4$He nuclear transitions~\cite{Krasznahorkay:2015iga,   Krasznahorkay:2018snd,Krasznahorkay:2021joi}. The excesses in both spectra can be interpreted as the production and subsequent decay into an $e^\pm$ pair of a new  boson, that was named 
$X_{17}$ after the fitted  value of its mass:
 \begin{eqnarray}
    \mxsv = \begin{cases}
    16.70 \pm 0.35 \pm 0.50  \ \textrm{MeV} \quad \qquad 
    ({}^8\textrm{Be~\cite{Krasznahorkay:2015iga}}) \\[0.3em]
           17.01 \pm 0.16  \ \textrm{MeV} \qquad \qquad  \qquad  ({}^8\textrm{Be~\cite{
           Krasznahorkay:2018snd}}) \\[0.3em]
        16.94 \pm 0.12 
        \pm 0.21 
        \  \textrm{MeV} \! \! \!  
   \qquad \quad \ \ ({}^4\textrm{He~\cite{Krasznahorkay:2021joi}}) \,,
\end{cases}
    \label{eq:m17}
\end{eqnarray}
where in the first and third result the first error is 
statistical and the second is systematic. 
In order to reproduce the strength of the observed excess a significant coupling to the quarks is required,
 along with a coupling to $e^\pm$ sufficiently large  
 to allow for the $\xsv \to e^+e^-$  decays to occur 
 within the ATOMKI apparatus. The nuclear data are not sufficient to fully determine the $\xsv$ spin/parity quantum numbers~\cite{Zhang:2020ukq,Feng:2020mbt,Viviani:2021stx}, 
 the most favourite possibilities being a vector or a pseudo-scalar  
 particle.

The dominant constraints on a particle with such couplings arise from the process $\pi^0 \to \gamma \xsv$ (followed by $ \xsv \to e^+e^-$) which has been thoroughly explored by the NA48/2 collaboration~\cite{NA482:2015wmo}.  These constraints imply that the  couplings of the $\xsv$  to matter 
must include a certain amount of 
pion-phobia~\cite{Feng:2016jff,Feng:2016ysn,Feng:2020mbt}. 
A large number of search strategies based primarily on the $\xsv$-quark interactions have been put forward recently, however, in general they tend to suffer from a large model dependence \cite{Ban:2020uii,Ellwanger:2016wfe,Alves:2017avw,Alves:2020xhf,Backens:2021qkv,Hostert:2020xku,Altmannshofer:2022izm}.
Due to its electron/positron interactions the $\xsv$ must abide with the standard search for ``visible'' dark photon, and strong  lower 
bounds on its coupling to electrons arise in particular from the  E141~\cite{Riordan:1987aw,Bjorken:2009mm,Andreas:2012mt,Liu:2017htz} and Orsay~\cite{Konaka:1986cb,Davier:1989wz} experiments, and more recently from the results of the NA64 collaboration~\cite{NA64:2019auh,NA64:2021aiq}. The visible search from the KLOE experiment~\cite{Anastasi:2015qla}
finally gives an upper limit at the per-mil level for the coupling to electrons. Covering the remaining allowed parameter space can provide a definite answer regarding the NP origin of the anomaly, and various experimental proposals 
could probe this region in the future~\cite{Balewski:2014pxa,Doria:2019sux,Baltzell:2022rpd}. 
Additional strong indirect bounds can be also obtained under mild theoretical assumptions. 
In particular, requiring that in the UV limit the $\xsv$ interactions respect weak $SU(2)$-invariance implies additional 
 constraints, in particular from neutrinos measurements,  see for instance Refs.~\cite{Altmannshofer:2022izm,Hati:2020fzp,Seto:2020jal,Nomura:2020kcw,Dutta:2020scq}.

This paper is devoted to describe a dedicated search strategy for a light bosonic state with mass around  $17\,$MeV, based on its production via \textit{resonant} annihilation of  positrons from the Frascati 
Beam Test Facility (BTF)~\cite{Ghigo:2003gy} beam on  atomic electrons in a fixed target $e^+ e^- \to \xsv$, with a subsequent prompt decay  $\xsv \to e^+ e^-$. The importance of the resonant production process, and the possibility of exploiting it  in  fixed target experiments that exploit  a primary positron beam, was first pointed out in Ref.~\cite{Nardi:2018cxi}.  The importance of this process was later recognised also for  electron and proton beam dump experiments, 
where positrons are produced in electromagnetic showers as secondary particles, and led to re-analysis of old results, 
 and improved projections for planned experiments~\cite{Marsicano:2018glj,Marsicano:2018krp,Celentano:2020vtu,Battaglieri:2021rwp,Andreev:2021fzd,Battaglieri:2022dcy}.

Our search strategy relies on an upgraded version of the PADME (Positron Annihilation into Dark Matter Experiment) experiment~\cite{Raggi:2014zpa,Raggi:2015gza} currently running at 
the Frascati National Laboratories (LNF).  
PADME exploits a positron beam from the DA$\Phi$NE LINAC accelerator in fixed target configuration with an active polycrystalline diamond target of $100\,\mu$m. We will discuss  in Sec.~\ref{sec:resonantprod} the $e^+ e^- \to \xsv $  production mechanism,
followed by prompt $\xsv \to e^+ e^-$ decay.  The planned   Run III of PADME that will be dedicated to search for the  $\xsv$ 
is described  in Sec.~\ref{sec:PADME}. A quantitative study of the expected sources of background  is presented in Sec.~\ref{sec:bkd}, and in Sec.~\ref{sec:proj} is used to infer projected limits.

\section{Resonant $\xsv$ production}
\label{sec:resonantprod}

Let us first consider the case where $\xsv$ is a spin-1  boson, which interacts with the Standard Model via the following Lagrangian:
\begin{eqnarray}
  \label{eq:LagrV}
\mathcal{L}^{\textrm{Vect.}} &\supset&   \sum_{f \ = \  e,u,d}  \xsv^{\mu} \, \bar{f}\gamma_{\mu} (\gv{f}+\gamma^5 \gvt{f}) f\,. 
\end{eqnarray}
All the relevant processes considered in this work are proportional to  the coupling combination  $(\gv{e}^2+\gvt{e}^2)$, up to terms involving the ratio $m_e^2/\mxsv^2< 10^{-3}$ which  will be neglected in the following. 
The dominant production mechanism for the $\xsv$ boson in the PADME experiment
(regardless of its parity and spin nature) is from the ``resonant'' process $e^+ e^- \to \xsv$.
For high-energy positrons impinging on the target electrons taken to be 
at rest,\footnote{This is a very good approximation for all electrons in the 2$s$, 2$p$ and 1$s$ 
atomic shells of $^{12}$C~\cite{Clementi:1963xxx,Nardi:2018cxi}.}
the resonant condition reads
\begin{equation}
\label{eq_PADMEreach_ResMass}
\Eres  = \frac{ \mxsv^2}{2 m_e} \ .
\end{equation}
We assume that the positron energies $E_+$ have a Gaussian  distribution with central value $E$ and spread $\sigma_E$\footnote{For positron energies  $E_+ \sim 10-20\,$MeV the typical energy loss  
in crossing a 100\,$\mu$m diamond target is of the order of $O(100)\,$keV. Typical values of the PADME beam spread 
$\sigma_E$ are of $O(1)$\,MeV, so distortion effects on the energy distribution can be neglected.}
\begin{align}
        f (E_+,E) = \frac{1}{\sqrt{2 \pi} \sigma_E } \displaystyle \ e^{ \displaystyle  -
        \frac{( E_{+} - E)^2}{2 \sigma_E^2}} \,.
\end{align}
We will work under the assumption of that the energy distribution of the positrons in the beam can be considered continuous when compared to the $\xsv$ width $\Gamma_X $, which implies the requirement:
\begin{align}
        \frac{\Gamma_X \mxsv}{2 m_e} \frac{1}{\sigma_E} N_{\rm tot} (E) \sim \left( \frac{N_{\rm tot} (E)}{1 \cdot 10^7} \right) \left( \frac{\gv{e}}{2\cdot 10^{-4}}\right)^2 \gg  1 \ ,
    \label{eq:continuouscondition}
\end{align}
where for simplicity of notations in the case of electron couplings we have defined $\gv{e} ~\equiv~ \left[\sqrt{\gv{f}^2 + \gvt{f}^2}\right]_{f=e}$.
In \eqn{eq:continuouscondition}
$N_{\rm tot} (E)$ is the total number of positrons in the beam with nominal energy $E$, and the strong inequality holds for $\sigma_E \sim O(1)\,$MeV.

At the leading order in QED, the resonant cross-section for production of a vector $X_{17}$ is given by:
\begin{equation}
\label{eq:resdelta1}
\sigma_{\rm res}^{\rm Vect.} = \frac{\gv{e}^2 \pi}{2 m_e}  \delta (E_+ - E_{\rm  res} ) \ .
\end{equation}

The final number of $\xsv$ produced  per positron-on-target for a given beam energy $E$ is thus given by
 \begin{align}
 \label{eq:nX17pot}
\mathcal{N}^{\,\rm per\ poT}_{\xsv} (E) =  \frac{\mathcal{N}_A Z \rho}{A} \ell_{\rm tar}  \frac{\gv{e}^2 \pi}{2 m_e} 
f ( \Eres , E)
\end{align}
where $\ell_{\rm tar}$ and  $\rho$ are respectively the target thickness and mass density.
For resonantly produced $\xsv$ the boost factor is rather small  $\gamma_{\xsv }  = \frac{ \mxsv}{2 m_e} \sim 17$, 
implying that even for 
the smallest experimentally allowed couplings,  
$\xsv$ decays will occur promptly, with  typical decay lengths  
never exceeding $O(1)\,$cm. 

The $\xsv$ particle mass is constrained by nuclear data
to lay in the limited range given in Eq.~\eqref{eq:m17}, 
which  suggests the range in which the beam energy
should be varied in order to optimise the $\xsv$ search strategy. 
 In this work we will consider a 
 ``conservative'' strategy  
 in which the beam energy is varied in the interval 
 $E\in [265,297]$ MeV, which 
 corresponds  to a scan in the centre-of-mass (CoM) energy range $\sqrt{s}\in [16.46,17.42]$ MeV,  
 that is a  $2\sigma$ range around the 
 $\xsv$ mass hint as measured in ${}^4$He.  
 We will also show the projected sensitivity for a more  
``aggressive'' 
search, in which  the beam energy 
 is restricted to vary in the interval $E\in [273,291]$ MeV,
 corresponding  to $\sqrt{s}\in [16.72,17.25]$ MeV, 
 that is a  $2\sigma$ range around the 
 $\xsv$ mass hint obtained by a naive combination 
 (i.e. neglecting possible correlations) of the ${}^4$He  and ${}^8$Be ATOMKI measurements~\cite{Krasznahorkay:2018snd,Krasznahorkay:2021joi}.

The sensitivity of the scanning procedure  depends on the energy step $\Delta E$ used in the scan.  For a Gaussian beam energy distribution 
the number of produced $\xsv$ falls exponentially fast when the 
mean beam energy $E$ departs from the resonance energy $\Eres$,
and reaches a minimum when $|E-\Eres|= \Delta E/2$. 
Denoting by $\alpha$ the relative variation of the projected limit 
 obtainable with the  highest vs. lowest production rates,
the energy step $\Delta E$ can be determined in terms of $\alpha$ as 
\begin{align}
\label{eq:energysteps}
    \Delta E \simeq 4 \sigma_E \sqrt{\alpha} \ .
\end{align}
For the projections of the PADME Run III sensitivity discussed below we require $\alpha \sim 20 \%$.
Finally, a useful approximation to the production rate
of a vector  $\xsv$, that in our setup with a  
$100 \mu\textrm{m}$ diamond target holds 
to a few percent level, 
is given by
\begin{equation}
\mathcal{N}^{\rm Vect.}_{X_{17}}
\simeq 1.8 \cdot 10^{-7}\ \times \left( \frac{\gv{e}}{2 \cdot 10^{-4}}\right)^2  \ \left( \frac{1 \ {\rm MeV} }{\sigma_E }\right) \ ,
\end{equation}    
where we have assumed that the beam energy is centred on the resonant energy $\sqrt{s} = M_X$.

In the case where the $\xsv$ is a pseudo-scalar particle (axion-like particle, or ALP in the following), the relevant Lagrangian is
\begin{equation}
  \label{eq:LagrALP}
\mathcal{L}^{\textrm{ALP}} ~\supset  \ \sum_{f \ = \ e,u,d}   \ga{f} m_f \, \xsv \, \bar{f} \gamma^5 f  \,.
\end{equation}    
%

The production cross-section is given by~\cite{Darme:2020sjf} 
\begin{equation}
\label{eq:resdelta}
\sigma_{\rm res}^{\rm ALP} = \frac{\pi  m_e \ga{e}^2}{4}
\times \, \delta (E_+ - E_{\rm  res} ) \,.
\end{equation}
Since photon couplings are independent of 
the initial and final state fermion chiralities, radiative correction are similar to the case of a vector $\xsv$. 
Thus, for the case of a pseudo-scalar $\xsv$ 
produced by a positron beam tuned at the resonant 
energy, and  decaying into an electron-positron pair,
we have
\begin{equation}
\label{eq:NperpotALP}
\mathcal{N}^{\textrm{ALP}}_{\xsv} \simeq 5.8 \cdot  10^{-7}\ \times \left( \frac{\ga{e}}{\textrm{GeV}^{-1}}\right)^2  \ \left( \frac{1 \ {\rm MeV} }{\sigma_E }\right) \, .
\end{equation}
In case the $\xsv$ decays also into photons or into other dark sector particles, this result should be multiplied by 
$\textrm{BR}(\xsv \to e^+e^-) $.


\section{The PADME experiment Run III}
\label{sec:PADME}

The DA$\Phi$NE  
BTF at LNF
offers interesting prospects for a $\xsv$ search based on resonant production~\cite{Nardi:2018cxi}. In particular, the LNF accelerator complex can provide a positron beam and vary its energy in the required range around 280 MeV. Assuming a typical diamond target (with electron density 
of $10^{24} \, \textrm{cm}^{-3}$) of $100\, \mu$m such as the one 
actually in use in the  
PADME experiment~\cite{Raggi:2014zpa,Raggi:2015gza} several thousand of $\xsv$ can be produced per $10^{10}$ positrons on target (PoT).

For this reason the PADME experiment has planned a dedicated data taking to search for $\xsv$ exploiting resonant production, 
called Run III.
For this run, starting in the autumn of 2022, the detector has been 
optimised to measure $\xsv$ visible decays. Using the ECal 
instead of the charged particle veto to detect the electron positron 
pairs will allow PADME to reach a much stronger rejection of the beam related background with respect to Run II conditions.
To suppress the photon background a new detector called ETag has 
been assembled. It is made of 5mm thick bars of plastic scintillators 
covering the front face of the PADME ECal. Due to the low Z of plastic 
scintillators and to their very small thickness, the bars will only be sensitive to charged particles, allowing to separate photons from electrons and positrons.

As mentioned above, PADME  plans to carry out a scan on various energy bins in order to cover thoroughly the interesting parameter space. Due to the fact that the $\xsv$ width is much smaller than the beam energy spread $\Gamma_X \ll \sigma_E$, 
the $\xsv$ particle is expected to contribute to the measured $e^+e^-\to e^+e^-$ rate in mostly a single bin of the scan, the one in which $|M_{17}-\sqrt s|$ is minimum. 
Since the excess will occur within a single energy point, the remaining ones will directly measure the background, so that the significance of the excess can be directly inferred from the data, without 
appealing to any MC simulation.

As  will be detailed in the next section, according to the predicted signal and background rates, that are summarised in 
Tab.~\ref{tab:BG_summary}, the excess 
of electron-positron pairs due to $\xsv$ production/decay could be at the sub percent level. 
Quantifying precisely the  confidence level of the 
exclusion limits (or  of a detected excess in the signal) will thus require an extremely precise control on the acquired luminosity point by point.
 Measuring simultaneously, using ECal, the rate for the $e^+e^-\to e^+e^-$ and for the $e^+e^-\to \gamma\gamma$ processes will allow to monitor the variation of their ratio which is not affected by the previous uncertainty. This strategy drastically reduces systematic effects with respect to 
 single measurements of the $e^+e^-\to e^+e^-$ rate, since systematic errors related to both, luminosity and acceptance measurements,  cancel  
 in the ratio.

\begin{figure}[h!]
\centering
        {\includegraphics[width=0.9\linewidth]{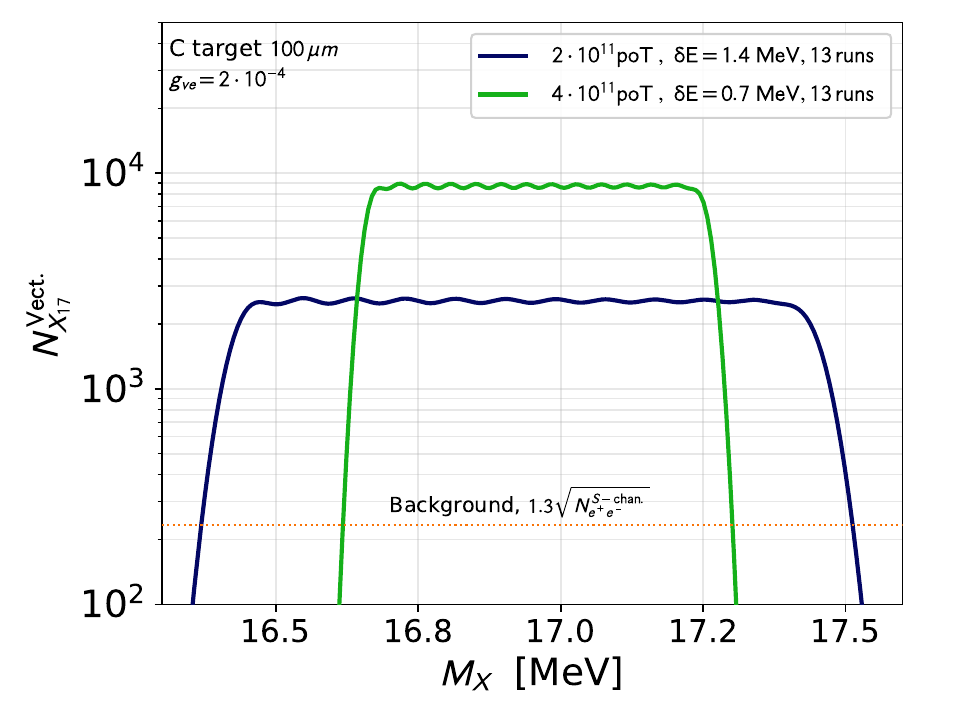}} \hspace{0.2cm}
        \caption{Number of expected vector $X_{17}$ as function of %
        $\mxsv$, for the conservative 
        (blue curve) and aggressive 
        (dashed green) scanning configurations for $\gv{e} = 2\cdot 10^{-4}$. 
        The dotted orange line corresponds to 
        the square root of the number of $e^+e^-$ events from $s$-channel off-shell photons, and illustrates the level of irreducible backgrounds.
        }
\label{fig:ScanNV}
\end{figure}

We will present the projection for the PADME Run III based on two scenarios for the total number of PoT and beam energy resolution:
\begin{itemize}
        \item \textit{Conservative}: 
        $2 \cdot 10^{11}$ total PoT on target, a $0.5\%$ beam spread, a broad energy range $[265,297]$, and an energy scan with $12$ bins.
   \item  \textit{Aggressive}:  
   $4 \cdot 10^{11}$ total PoT on target, a $0.25\%$ beam spread, a narrower energy range $[273,291]$, and an energy scan with $14$ bins.
\end{itemize}
In both cases, the number of steps were optimised based on the averaged projected limits. Note that the reach is only midly reduced by further increasing the number of steps. Both scanning  
strategies are illustrated in Fig.~\ref{fig:ScanNV}, where a first estimate of the irreducible background level is also shown.  The total number of positron on target  per energy point required is of the order of 
$10^{10}$.
Using $\sim$ 2500 positron on target per bunch PADME will be able to collected the necessary statistics in  a few days of fully efficient running.
Reducing the beam spread $\sigma_E$ implies that 
a larger number of energy steps must be used to cover 
the whole interesting region. 
For a fixed number of total positrons on target, the larger production rate per positron from Eq.~\eqref{eq:nX17pot} is therefore compensated by the smaller number of positrons available for each energy point.
However, as the number of background events is proportional to the number
of impinging $e^+$, reducing the energy beam spread ultimately increases 
the signal-to-noise ratio in each energy bin, thus rising the sensitivity
to signatures of NP  and improving the projected experimental reach.

\section{Main 
background processes}
\label{sec:bkd}
The use of resonant production offers a unique opportunity for enhancing the $\xsv$ productions rates compared to the 
associated  $e^+e^- \to \gamma \xsv$ production. However, it also introduces challenging QED background sources which are hard to constrain.
Given that the resonant production of $\xsv$ requires a low 
CoM energy of $\sim 17\,$MeV, the main background sources are:
\begin{itemize}
    \item $t$- and $s$-channel  Bhabha scattering 
    \item $e^+e^- \to \gamma\gamma$
    \item $e^+N \to e^+N+\gamma$
\end{itemize}
While Bhabha scattering is producing the same final state as  $\xsv$ decays, the remaining processes produce at least one photon. In detectors using a pure calorimetric approach photons and electrons are indistinguishable,  implying that a  non negligible background contribution from photons final state can arise.
We will first concentrate  on Bhabha scattering, assuming that photons background can be controlled by identifying photons in the final state, as will be discussed in the Sec.~\ref{sec:photonbkd}.
We will study the two Bhabha contributions separately to provide a better understanding of the physics at fixed target, neglecting their interference term. The interference term has been checked to be negative and to produce a reduction of the total cross section below the \% level. 

\subsection{The Bhabha scattering background}

\begin{figure}[h!]
\centering
        {\includegraphics[width=0.9\linewidth]{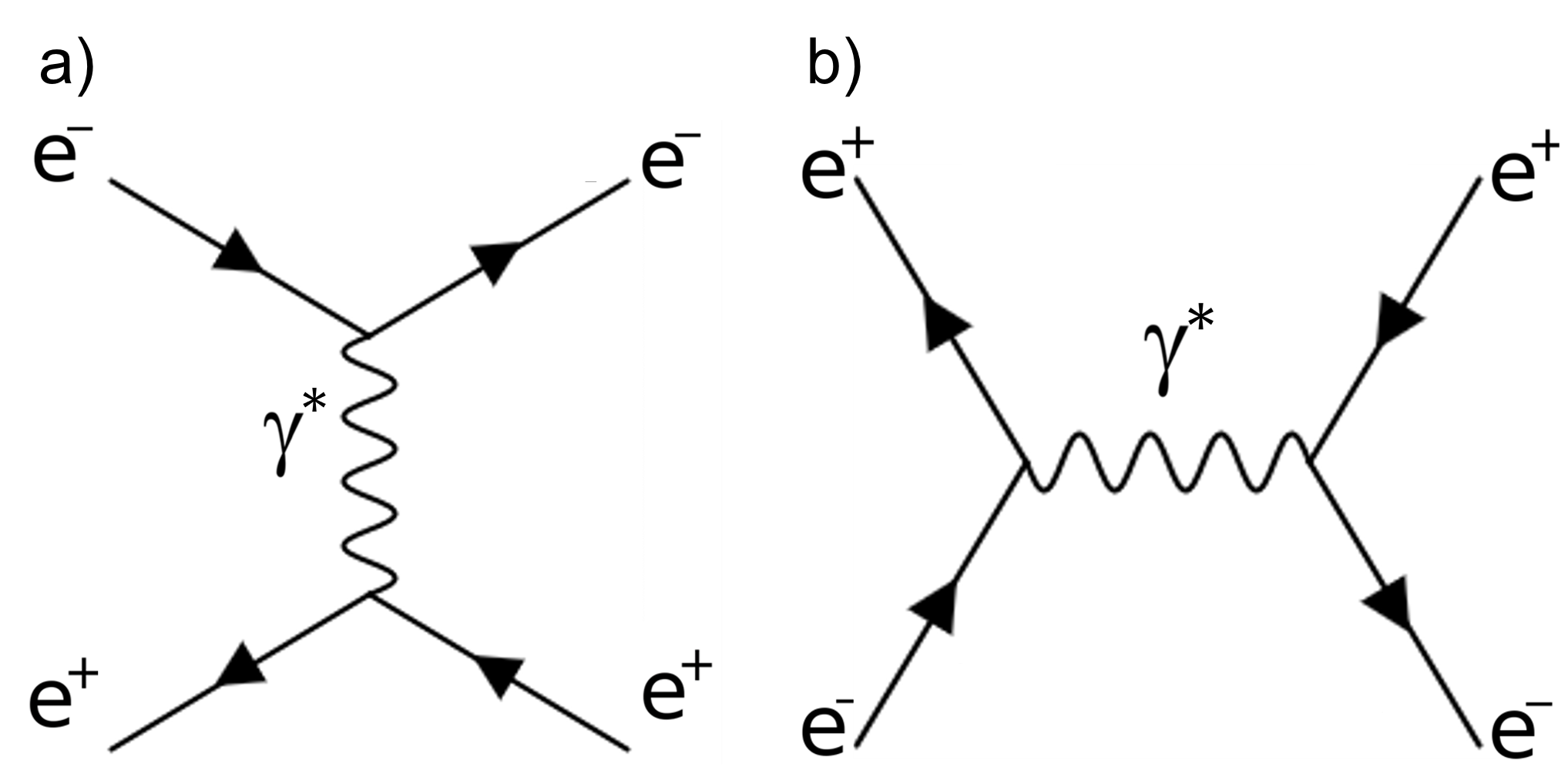} }
        \caption{ a)  $t$-channel and  b) $s$-channel diagrams contributing to  Bhabha scattering in the SM.
        }
\label{fig:Bhabha_Diag}
\end{figure}

\begin{figure}[h!]
\centering
        {\includegraphics[width=0.9\linewidth]{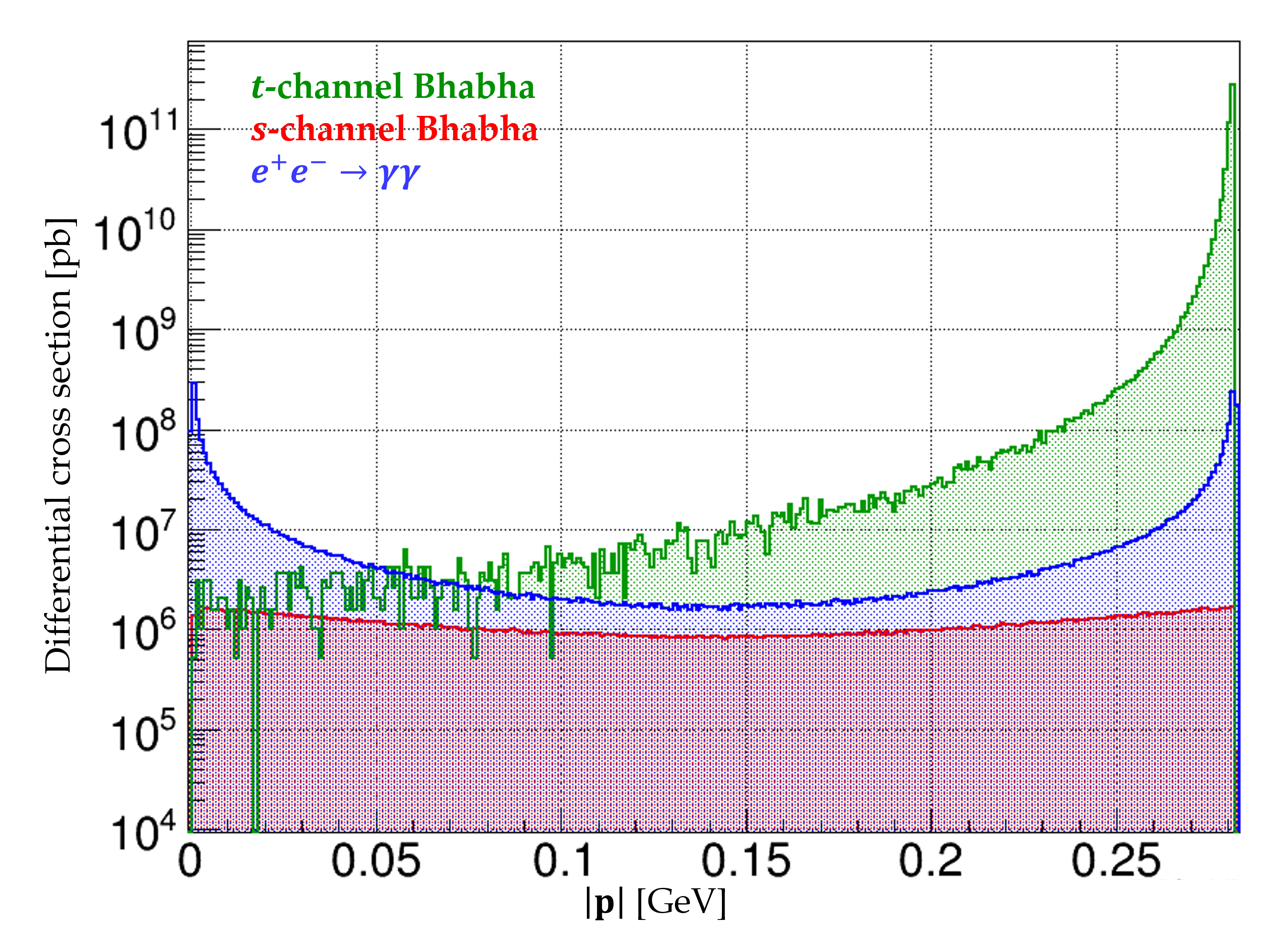} }
        \caption{Differential cross sections for the main sources of background as a function of         
        the three-momentum of the outgoing positron (resp. photon)
        $|\mathbf{p}|$.}
       
\label{fig:eeDiffCross}
\end{figure}

The unique Standard Model (SM) process that can produce final states identical to the $\xsv$ decays at CoM energies in the MeV range is $e^+e^- \to \gamma^* \to e^+e^-$, where $\gamma^*$ denotes an off-shell photon. 
In the SM,  Bhabha scattering proceeds via two different contributions, namely the $t$-channel and $s$-channel 
amplitudes depicted in Fig.~\ref{fig:Bhabha_Diag}. 
While the final state is the same, %
their kinematic, especially in in fixed target experiments, is very different. The $t$-channel process, despite being dominant in terms of contributions to the total cross section, can be efficiently rejected having a quasi elastic behaviour, with the positron retaining almost all of its original energy, and the electron remaining almost at rest (green distribution in Fig.~\ref{fig:eeDiffCross}). This is a characteristic of the fixed target experiments in which the initial energy of the two particles is very different. Even exploiting the different kinematics the contribution of $t$-channel process 
cannot be neglected.
The $s$-channel process on the other hand has the same kinematics of the signal $e^+e^- \to \xsv \to e^+e^-$ because the virtuality of the off-shell  photon is $\sqrt{s}\sim M_X$. 
For this reason the $s$-channel process constitutes an irreducible  source of background, and represents the limiting factor for the sensitivity of experiments that exploit resonant production.

\subsection{Photon background}
\label{sec:photonbkd}
In order to control the photon related background 
a detector separating electron from photon clusters is required.
For this reason during Run III a new detector called ETag will be introduced in the original PADME setup. ETag will produce signals only if the crossing particle is electrically charged, with a few percent mis-tagging probability of identifying a photon as a charged particle. 

The two photon annihilation process is the most relevant photon based  source of background due to the large cross section and reduced effectiveness of the kinematical constraints.
Assuming a mis-tagging probability $\epsilon_{mis}=5\%$ a rejection factor on the $e^+e^- \to \gamma\gamma$ process of $1/\epsilon_{mis}^2=400$ can be obtained. Such a value, 
if achievable,  would reduce the background originating from two photon annihilation to a negligible level.

Positron bremsstrahlung can also originate events with two charged particles, if the radiated photon is mis-tagged as a charged particle, and the two particles are entering the ECal acceptance. Both these conditions are rare, being the process dominated by high energy forward positrons and soft forward photons crossing the PADME ECal central hole. In addition, the two cluster eventually produced in the ECal will hardly reproduce the correct invariant mass value of 17 MeV. For this reasons the background coming from bremsstrahlung can be considered as negligible. 

\subsection{Beam related and pile up backgrounds}

\begin{table}[t!]
    \centering
    \begin{tabular}{l|c|c|c}
    BG Process & \# of Ev. & \# of Ev. in Acc. & Acc.\\
    \hline     
    $e^+e^- \to e^+e^-$ ($t\,$-\,ch.) & $5.4 \cdot 10^7$ & $6.9\cdot 10^4$ & $0.13$\% \\
    $e^+e^- \to e^+e^-$ ($s$\,-\,ch.) &$3.2\cdot 10^4$ &  $6.4 \cdot 10^3$ & $20$\%\\
    $e^+e^- \to \gamma\gamma$  & $2.9\cdot 10^5$ &$1.3\cdot 10^4$ & $4.5$\%\\
    \hline
    $e^+e^- \to \xsv \to e^+e^-$ & $1250$ & $250$ & 20\%\\
    \end{tabular}
    \caption{Expected number of background and signal events per $1\times10^{10}$ positrons on target. The $t-$channel values before selection cuts correspond to $e^\pm$ with energies larger than $1\,$MeV. 
    The acceptance cuts do not include the $\gamma \gamma$ tagging from the ETag.
    }
    \label{tab:BG_summary}
\end{table}

An additional possible source of background comes from overlapping beam-target interactions or beam related background.
In the first case interactions of two different primary positrons can produce two simultaneous clusters in the PADME ECal. In the second case the interactions of the beam halo with beam line elements can produce multiple clusters events. These sources of background are proportional to the beam intensity which will be reduced to few thousand $e^+$/bunch in PADME Run III, a factor 10 lower with respect to PADME Run II.

Combinatorial backgrounds sources can be controlled by applying several kinematic constraints to the event selection. The Bhabha kinematics is highly constrained so that the energy and the polar angle of each of the leptons are connected by an analytic expression,  $E_{e^\pm}=f(\theta_{e^\pm})$. A simple set of conditions can be applied to reject combinatorial backgrounds:
\begin{itemize}
    \item $E_{e^+}+E_{e^-}=E_{beam}$
    \item $M^2(e^+e^-)= \sqrt{s} $ 
    \item $E_{e^\pm}=f(\theta_{e^\pm})$
\end{itemize}
This strategy is well understood, and has been proven to be 
quite effective  in the 
analysis of $e^+e^- \to \gamma\gamma$ events during PADME Run
II~\cite{GammaGammaPADME}. 

\subsection{Expected backgrounds summary}

Let us now summarise the expected background contribution obtained by simulating final state kinematics with
the \calchep~package. We compute the total number of expected events using the cross section provided by \calchep, and we evaluate the acceptance adopting the following strategy:
\begin{itemize}
    \item The energy of both outgoing particles $E_1 ,E_2$ is in the range: $ E_{1,2}>$ 100 MeV.
    \item The azimuthal angle of both particles $ \theta_{1}, \theta_{2}$ is in the range:  25.5 $\lesssim \theta_{1,2}/{\rm mrad} $ $\lesssim$\,77.
\end{itemize}
Applying these conditions we find the number of signal and background 
events for each energy point of the scan ($\sim 1\times10^{10}$ poT). Our  results are summarised in Tab.~\ref{tab:BG_summary}. 
The signal rate has been obtained for $\gv{e}=2\cdot 10^{-4}$ which saturates the unexplored region of parameter space for vector particles. For the above background level, projected limits at $90\%$ C.L. will thus correspond to having less than $\sim 360$ NP events.\footnote{We assume that the $\gamma \gamma$ background will be reduced to a negligible level from the tagging in the ETag detector.}
Tab. \ref{tab:BG_summary} shows that     the $\xsv$ production rate is non negligible with respect to the background even for small values of the $\gv{e}$ coupling. The resulting signal acceptance has been obtained using Bhabha $s$-channel kinematics, which is expected to be identical to the $\xsv$ one for prompt $\xsv$ decays. The actual acceptance value will depend on the final experimental cut strategy.

\section{Projections}
\label{sec:proj}

\begin{figure}[!t]
\resizebox{0.49\textwidth}{!}{%
  \includegraphics{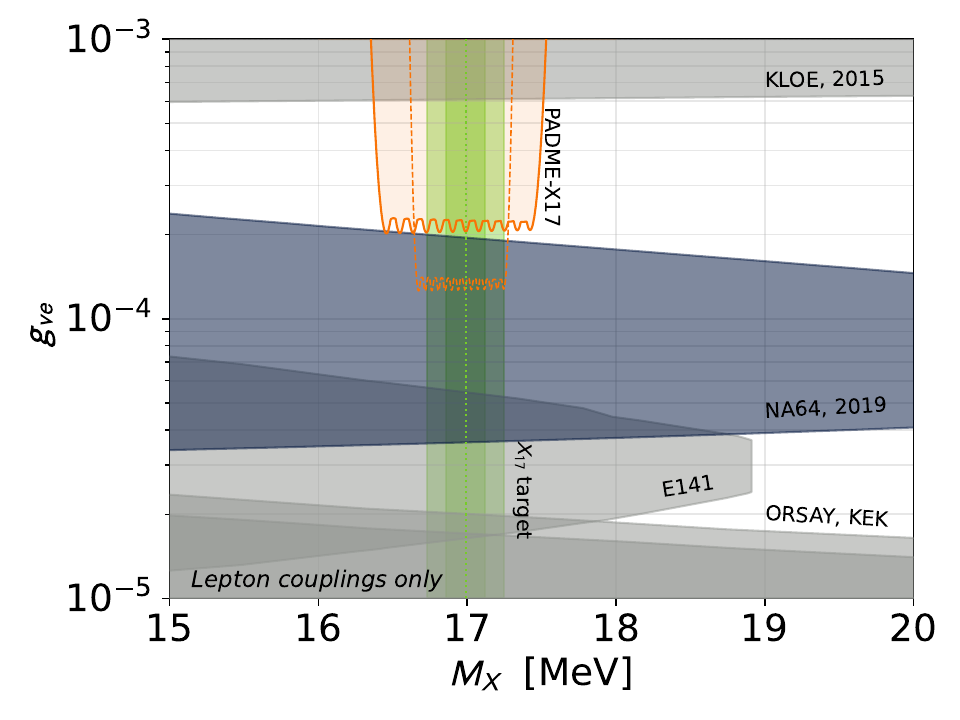}
}
\caption{Projected $90 \%$ C.L. sensitivity of PADME Run-III on the  $g_{ve}$  coupling of a $\xsv$  vector boson for the conservative (solid orange line) and aggressive (dashed orange line) setups. 
 Lepton-based experimental limits 
 from the KLOE~\cite{Anastasi:2015qla}, NA64~\cite{NA64:2021aiq},
 E141~\cite{Riordan:1987aw}, KEK and Orsay~\cite{Konaka:1986cb,Davier:1989wz} experiments are also shown. The dark (light) green band represents the $1\sigma$  ($2\sigma$) $\xsv$ mass target from a naive combination of the ${}^4\textrm{He}$ and ${}^8\textrm{Be}$ ATOMKI  results~\cite{Krasznahorkay:2018snd,Krasznahorkay:2021joi}.}
\label{fig:X17leptons}       
\end{figure}

\begin{figure}[!h]
\resizebox{0.49\textwidth}{!}{%
  \includegraphics{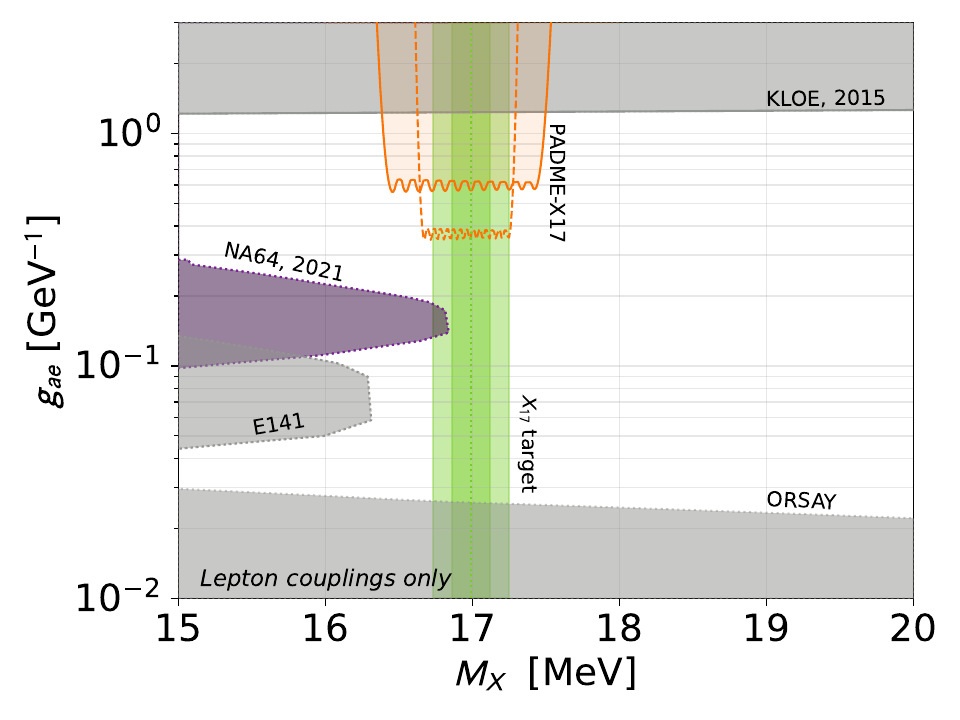}
}
\caption{Projected $90 \%$ C.L. sensitivity of PADME Run-III on the  $g_{ae}$  coupling of a  $\xsv$ ALP for the conservative (solid orange line) and aggressive (dashed orange line) setups. Lepton-based experimental limits 
from the KLOE~\cite{Anastasi:2015qla}, NA64~\cite{NA64:2021aiq}, 
E141~\cite{Riordan:1987aw} and  Orsay~\cite{Davier:1989wz} experiments are also shown. 
The dark (light) green band represents the $1\sigma$  ($2\sigma$) $\xsv$ mass target from a naive combination of  ${}^4\textrm{He}$ and ${}^8\textrm{Be}$ ATOMKI results~\cite{Krasznahorkay:2018snd,Krasznahorkay:2021joi}.}
\label{fig:X17leptons_ALP}       
\end{figure}
In  Fig~\ref{fig:X17leptons}
we show the projections of the constraints on $g_{ve}$ 
for the case of vector $\xsv$ as function 
of its  mass.  
For a vector particle with mass $M_X\approx 17\,$MeV 
the parameter space below  $g_{ve} \approx 0.2 \cdot 10^{-3}$ is excluded by a combination of E141~\cite{Riordan:1987aw} and NA64 experiments~\cite{NA64:2019auh} as well as by other beam dump experiments~\cite{Konaka:1986cb,Davier:1989wz}. As can be seen from the picture, in the vector boson case the projected sensitivity of PADME Run-III reaches down all the way to the upper limit from NA64, completely covering the still viable parameter space region. PADME Run-III will thus exhaustively probe the hypothesis that the ATOMKI anomaly is due to a new vector boson with mass  $M_X\approx 17\,$MeV.

In the case the $\xsv$ is instead a spin-0 ALP, the existing limits are significantly less constraining due to the somewhat  shorter lifetime and the slightly reduced production rate of an $\xsv$ ALP 
compared to a 
vector boson. 
In particular, we see from Fig.~\ref{fig:X17leptons_ALP} 
that the NA64 limit~\cite{NA64:2019auh} is barely reaching into  
 the parameter space region favoured by the nuclear experimental  data.
 In contrast, the  projections corresponding to the regions in 
 orange 
indicate that PADME Run-III will be able to probe a significant part of the viable parameter space for an $\xsv$ ALP.

\section{Conclusion}

 In this work we have described a novel approach to probe the existence of the putative $\xsv$ boson hinted by the  
 anomalies in $^8$Be and $^4$He nuclear transitions
reported by the ATOMKI collaboration, 
 and more generally of any new light boson 
 with a mass close to $17$ MeV and coupled to $e^\pm$. By exploiting  the resonant annihilation of positrons of an energy-tuned  beam on  atomic electrons in a thin target, the scan-based procedure 
 that we have described can be used to extract the signal, while fitting  the background directly from the off-resonance data.

We have studied the implementation 
of this strategy in the upcoming Run-III of the PADME experiment 
at LNF. We have worked out a number of projections based on a statistic of a few$\,\times 10^{11}$ total number of positrons on target. 
We have considered only statistical errors on signal and background. 
However, there are good reasons to expect that 
systematic errors could be kept to a similar level, including in particular the uncertainty on the relative number of positrons on target collected at each energy point of the scan.
In this case 
several weeks of data-taking at  PADME  would be sufficient  to  cover completely the  parameter space region  still viable for  a spin-1 $\xsv$ candidate. In contrast, 
it will not be possible to completely exclude (or discover with certainty)  a spin-0  $\xsv$ ALP, but still it will be possible to reduce significantly the viable parameter space.   
It is  worth mentioning at this point that, since the remaining  parameter space for an $\xsv$ ALP lays is the small coupling region, 
typical decay lengths can be as large as $\sim 1\,$cm. Hence, if a sufficiently precise vertexing of the decay could be  engineered,  the background would be dramatically reduced, and  eventually it might be  possible to close also this region.

While this work has focused on the mass range 
$M_X \in  [16 , 18]\,$MeV as determined by the BTF beam energy range, 
 the same technique could be used in the future to probe all
 types of light  bosons feebly interacting  with $e^\pm$ with masses in the tens of MeV range, based on the availability of positron  beams of adequate and tunable energies. 

\subsection*{Note added}
Shortly after the completion of this paper an arXiv   preprint of the ATOMKI collaboration appeared~\cite{Krasznahorkay:2022pxs} 
that reports the observation of a further  anomaly
in the large angles correlation 
of  $e^+e^-$  pairs produced in  $^{12}$C nuclear transitions. 
This new anomaly is consistent with the $\xsv$ vector boson interpretation of the  $^8$Be and $^4$He anomalies, but it is at odd with a $\xsv$ of pseudoscalar  nature~\cite{Feng:2020mbt}.

\subsection*{Acknowledgements}
\noindent

We acknowledge several discussions with members of the PADME collaboration 
and in particular P. Valente for fundamental inputs on the LNF beam setup, and P. Gianotti 
for discussions on the details of the PADME experimental strategy. 
L.D. and E.N.~have been supported in part by the INFN ``Iniziativa Specifica'' Theoretical 
Astroparticle Physics (TAsP-LNF). E.N. and M.R. are partially supported by the Sapienza Grant ``Ricerca del bosone X17 nell'esperimento PADME ai laboratori Nazionali di Frascati.''. L.D. is  supported by the European Union’s Horizon 2020 research and innovation programme under the Marie Skłodowska-Curie grant agreement No 101028626 from 01.09.2021. M.M. acknowledges professor A. D'Angelo for bringing him in contact with Beyond Standard Model Physics.

\bibliographystyle{utphys}
\bibliography{PADMEX17}

\end{document}